\documentstyle{article}
\font\cero=cmss10 scaled 1728
\font\uno=cmssbx10 scaled 1200
\begin{document}
\begin{flushleft}
{\cero Adjoint operators and perturbation theory of black holes} \\[3em]
\end{flushleft}
{\sf R. Cartas-Fuentevilla}\\
{\it Instituto de F\'{\i}sica, Universidad Aut\'onoma de Puebla,
Ap. postal J-48 72570, Puebla Pue., M\'exico, and Enrico Fermi Institute,
University
of Chicago, 5640 S. Ellis Ave., Chicago, Illinois 60637} \\ [4em]

We present a new approach for finding conservation laws in the perturbation
theory of black holes which applies for the more general cases of
non-hermitian equations governing the perturbations. The approach
is based on a general result which establishes that a covariantly conserved
current can be obtained from a solution of any system of homogeneous linear
differential equations and a solution of the adjoint system. It is shown
that the results obtained from the present approach become essentially the
same (with some differences) to those obtained by means of the traditional
methods in the simplest black hole geometry corresponding to the
Schwarzschild spacetime. The future applications of our approach
for studying the perturbations of black hole spacetimes in string theory 
is discussed.
\vfill

\noindent Running title: Adjoint operators and....
\newpage

\begin{center}
{\uno I. INTRODUCTION}
\end{center}
\vspace{1em}

The problem we consider here is in many ways an old one, the scattering of
test fields by a black hole background. Several methods for solving this
problem have been applied, wave mechanical scattering, semiclassical
methods, etc. These methods treat the scattering of massless waves from
black holes in a manner completely analogous to the Schr\"odinger scattering
of matter waves in an effective potential. More specifically, in the scheme
of the classical Einstein-Maxwell theory, after separation of harmonic
time and angular dependence, the coupled gravitational and electromagnetic
perturbations of the Reissner-Nordstr\"om black hole (which contains the
perturbations of the Schwarzschild black hole as a special case) can be
described in terms of solutions of a Schr\"odinger-type equation (see
\cite{1} and references therein). The special properties of such an
equation are widely known, but we shall discuss them briefly for clarifying
our aims. Before all, this equation is Hermitian (self-adjoint), which is
directly connected with the constancy of the Wronskian of two solutions,
which in turn, permits to establish a conservation relation between
incoming energy flux from infinity, the flux back out to infinity, and the
energy crossing the horizon of the black hole (see \cite{1} and references
therein). However, when one considers the general (coupled) perturbations
of some solutions in the framework of the modern theories such as string
theories involving nonminimally coupled scalar fields, the field
perturbations appear to be governed by non-hermitian systems \cite{2}, and
in this manner the above ideas for hermitian systems are no longer applied.
The purpose of this paper is to demonstrate that similar conservation
relations can be obtained for any non-hermitian equation systems using the
concept of the adjoint of a linear operator; the hermitian case becomes only
a particular one. These results are of interest for understanding scattering
phenomena in the setting of modern theories, and possibly in other areas of
physics. However, in the present work we shall apply our approach to the
simplest black hole spacetime, the Schwarzschild solution, mainly for two
reasons. First, for demonstrating that the present approach gives
essentially the same very known results (however, see \cite{1}) in the
perturbation theory for that solution. Second, for clarifying our basic
ideas in the simplest case.

The plan of this paper is as follows. Section II presents the general
relationship between adjoint operators and covariantly conserved currents.
The corresponding currents for the gravitational and electromagnetic
perturbations of the Schwarzschild solution and their separations of
harmonic time and angular dependence are given also in this section. The
asymptotic solutions at the spatial infinity and at the horizon for the
non-hermitian equations governing the perturbations are discussed in Sec.\
III. In Sec.\ IV we calculate the field perturbations, fluxes of energy,
and using the results of the previous sections, the conservation relations
are established. Furthermore, the energy reflection and transmission
coefficients are discussed. Finally, in Sec.\ V we discuss the conclusions
and the future applications of our present approach. \\[2em]

\noindent {\uno II. ADJOINT OPERATORS AND CONSERVED CURRENTS}
\vspace{1em} \\

In Refs.\ \cite{3,4} it has been shown that there exists a conserved
current associated with any system of homogeneous linear partial
differential equations that can be written in terms of a self-adjoint
operator. This result is limited for a self-adjoint system, for which the
corresponding conserved current depends on a pair of solutions admitted by
such a system. However, as we shall see below, there exists a more general
possibility that extends for systems of equations which are not self-adjoint
necessarily. The demonstration is very easy, following the basic idea of
Refs.\ \cite{3,4}:

In accordance with Wald's definition \cite{5,6}, if ${\cal E}$ corresponds
to a linear partial differential operator [such as the non-hermitian
operators governing the potentials $\psi_{\rm G}$ and $\psi_{\rm E}$ in
Eqs.\ (2) and (7) below] which maps $m$-index tensor fields into $n$-index
tensor fields, then, the adjoint operator of $\cal E$, denoted by ${\cal
E}^{\dag}$, is that linear partial differential operator mapping $n$-index
tensor fields into $m$-index tensor fields such that
\begin{equation}
     g^{\rho\sigma\ldots} [{\cal E}(f_{\mu\nu\ldots})]_{\rho\sigma\ldots} -
     [{\cal E}^{\dag}(g^{\rho\sigma\ldots})]^{\mu\nu\ldots} f_{\mu\nu\ldots}
     = \nabla_{\mu}  J^{\mu},
\end{equation}
where $J^{\mu}$ is some vector field.

From Eq.\ (1) we can see that this definition automatically guarantees that,
if the field $f$ is a solution of the linear system ${\cal E} (f) = 0$ and
$g$ a solution of the adjoint system ${\cal E}^{\dag} (g) = 0$, then
$J^{\mu}$ is a covariantly conserved current (which depends on the fields
$f$ and $g$). This fact means that for any homogeneous equation system, one
can always to construct a conserved current taking into account the adjoint
system. This general result contains the self-adjoint case as a particular
one. The physical meaning of such a conserved current will depend on each
particular case.

With this result, our purpose is to reach the same physical conclusions
reached in Ref.\ \cite{1}, starting directly from non-hermitian Eqs.\ (2)
and (14) (without invoking the hermitian equivalent system (21) of that
reference). These equations represent the type of systems of equations that
one can obtain for describing the perturbations of the exact solutions in
the modern theories, but that, however, do not appear to take the form of a
hermitian system.

In the following, the formulae and notation of Ref.\ \cite{1} will be used
extensively.
\vspace{1em}
\begin{center}
{\uno A. Gravitational perturbations}
\end{center}
As it is well known, gravitational perturbations of Schwarzschild solution
can be described in terms of the scalar potential $\psi_{\rm G}$ satisfying
\begin{equation}
     {\cal O}^{\dag}_{\rm G} (\psi_{\rm G})= [ (\Delta + 2 \gamma + \mu) (D
     + 3 \rho) - (\overline{\delta} - 2 \overline{\beta}) (\delta + 4 \beta)
     - 3 \Psi_{2} ] \psi_{\rm G} = 0,
\end{equation}
which can be obtained from Eqs.\ (2) and (4) of Ref.\ \cite{1}. Equation\ (2)
corresponds to the adjoint system of the decoupled equation satisfied by the
perturbed Weyl spinor $\Psi^{\rm B}_{0}$ according to Wald's method \cite{5}:
\begin{equation}
     {\cal O}_{\rm G} (\Psi_{0}^{\rm B})= [ (D - 5 \rho) (\Delta - 4 
\gamma
     + \mu) - (\delta - 2 \beta) (\overline{\delta} + 4 \overline{\beta})
     - 3 \Psi_{2} ] \Psi^{\rm B}_{0} = 0,
\end{equation}
in this manner, from Eqs.\ (1)--(3) and considering that the tetrad
components $D$, $\Delta$, $\delta$, and $\overline{\delta}$ are defined by
$l^{\mu} \partial_{\mu}$, $n^{\mu} \partial_{\mu}$, $m^{\mu}
\partial_{\mu}$, $\overline{m}^{\mu} \partial_{\mu}$, respectively, and that
they act on scalar fields we have the following continuity equation
\begin{equation}
      \nabla_{\mu} \big[ \Psi^{\rm B}_{0}[n^{\mu}(D + 3\rho) \psi_{\rm G} -
      \overline{m}^{\mu} (\delta + 4\beta) \psi_{\rm G}] + \psi_{\rm G} [
      m^{\mu} (\overline{\delta} + 4\overline{\beta}) \Psi^{\rm B}_{0} -
      l^{\mu} (\Delta - 4\gamma + \mu) \Psi^{\rm B}_{0}] \big] = 0.
\end{equation}

In order to separate the time and angular dependence, we have consider the
expressions (6) and (41) of Ref.\ \cite{1} for $\psi_{\rm G}$ and $\Psi^{\rm
B}_{0}$ in terms of the harmonic time and the spin-weighted spherical
harmonics, and we obtain
\begin{equation}
      \partial_{r} \big[ \chi (\Psi^{\rm B}_{0} \partial_{r} \psi_{\rm g}
      - \psi_{\rm g} \partial_{r} \Psi^{\rm B}_{0}) -  2 (\partial_{r} \chi)
      \psi_{\rm g} \Psi^{\rm B}_{0} \big] = 0,
\end{equation}
where $\chi = r^{2} - 2 M r$, and we have used the expressions\ (3a) and (4)
of Ref.\ \cite{1} for the background spin coefficients and the tetrad and
preserved the same symbol for $\Psi^{\rm B}_{0}$ as describing its radial
dependence, $\psi_{\rm g}$ is the corresponding one for $\psi_{\rm G}$.
Therefore, Eq.\ (5) implies that there exists a conserved quantity which
will be denoted by $K_{\rm G}$:
\begin{equation}
      \chi (\Psi^{\rm B}_{0} \partial_{r} \psi_{\rm g} - \psi_{\rm g}
      \partial_{r} \Psi^{\rm B}_{0}) -  2 (\partial_{r} \chi)
      \psi_{\rm g} \Psi^{\rm B}_{0} \equiv K_{\rm G}.
\end{equation}
\begin{center}
{\uno B. Electromagnetic perturbations}
\end{center}
Similarly, the electromagnetic perturbations are described in terms of the
scalar potential $\psi_{\rm E}$ which satisfies \cite{1}
\begin{equation}
     {\cal O}^{\dag}_{\rm E} (\psi_{\rm E})= [ (D + \rho) (\Delta + 2 \gamma
     + \mu) - (\delta + 4 \beta) (\overline{\delta} - 2 \overline{\beta})
     - 6 \Psi_{2} ] \psi_{\rm E} = 0,
\end{equation}
and its adjoint system for the component of perturbed electromagnetic field
$\varphi^{\rm B}_{0}$ :
\begin{equation}
     {\cal O}_{\rm E} (\varphi^{\rm B}_{0}) = [ (\Delta - 4 \gamma + \mu)
     (D - 3 \rho) - (\overline{\delta} + 4 \overline{\beta}) (\delta - 2
     \beta) - 6 \Psi_{2} ] \varphi^{\rm B}_{0} = 0,
\end{equation}
and following the same procedure employed in the gravitational case we
obtain the corresponding continuity equation:
\begin{equation}
      \nabla_{\mu} \big[ \varphi^{\rm B}_{0} [ l^{\mu}(\Delta + 2 \gamma +
      \mu) \psi_{\rm E} - m^{\mu} (\overline{\delta} - 2 \overline{\beta})
      \psi_{\rm E}] + \psi_{\rm E} [\overline{m}^{\mu} (\delta - 2 \beta)
      \varphi^{\rm B}_{0} - n^{\mu} (D - 3 \rho) \varphi^{\rm B}_{0}] \big]
      = 0.
\end{equation}
After the separation of the time and angular dependences (see Eqs.\ (15) and
(42) of Ref.\ \cite{1}), Eq.\ (9) yields a conserved quantity for
electromagnetic perturbations:
\begin{equation}
      \big[ \chi (\psi_{\rm e} \partial_{r} \varphi^{\rm B}_{0}
      - \varphi^{\rm B}_{0} \partial_{r} \psi_{\rm e}) + (\partial_{r} \chi)
      \psi_{\rm e} \varphi^{\rm B}_{0} \big] \equiv K_{\rm E},
\end{equation}
where $\varphi^{\rm B}_{0}$ has been preserved as describing its radial
dependence and $\psi_{\rm e}$ corresponds to that for $\psi_{\rm E}$. Since
$\Psi^{\rm B}_{0}$ appearing in Eq.\ (4) is invariant under the ordinary
gauge transformations of the metric perturbations \cite{1}, the current
itself has the same invariance property; something similar occurs with
$\varphi^{\rm B}_{0}$ appearing in Eq.\ (9) and the corresponding current,
as opposed to the currents constructed by means of the approaches of
Refs.\
\cite{7,8} and Refs.\ \cite{3,4}. Similarly $\Psi^{\rm B}_{0}$ and
$\varphi^{\rm B}_{0}$ are independent on the perturbed tetrad gauge freedom
(see \cite{9} are references therein). These properties are important when
$K_{\rm G}$ and $K_{\rm E}$ are evaluated.

On the other hand, the covariantly conserved currents given in Eqs.\ (4) and
(9) depend on the background field and on the solutions admitted by the
corresponding original system and its adjoint system. However, by noting
that $\Psi^{\rm B}_{0} \: (\varphi^{\rm B}_{0})$ in Eqs.\ (4)--(6) (Eqs.\
(9) and (10)) is defined in terms of derivatives of $\overline{\psi}_{\rm G}
\: (\overline{\psi}_{\rm E})$ according to Eqs.\ (41) and (42) of Ref.\
\cite{1}, our conserved currents depend actually on a single complex
solution of the corresponding equation for the potential.

For establishing our conservation relation for the field perturbations, the
idea is to evaluate the constants $K_{\rm G}$ and $K_{\rm E}$ at the two
asymptotic regions, the spatial infinity and the horizon, and then it will
be clarified the physical meaning of these constants. With this purpose, in
the next section the asymptotic forms of the potentials $\psi_{\rm G}$ and
$\psi_{\rm E}$ at those regions will be determined, since all the quantities
of interest are defined in terms of them.  \\[2em]

\noindent {\uno III. ASYMPTOTIC EXPANSIONS}
\vspace{1em}
\begin{center}
{\uno A. At the spatial infinity}
\end{center}
Since $\psi_{\rm G}$ has spin weight --2, we seek solutions of Eq.\ (2) of
the form \cite{1}
\begin{equation}
     \psi_{\rm G} = \psi_{\rm g} (r) e^{-i\omega t} {_{-2}}Y_{jm} (\theta
     ,\varphi),
\end{equation}
after substituting (11) into Eq.\ (2), one gets a linear ordinary
differential equation for $\psi_{\rm g} (r)$. Such a system has essentially
the same form of Eq.\ (2) just making the substitutions  $\Delta$, $D$, and
$(\overline{\delta} - 2 \overline{\beta}) (\delta + 4 \beta)$ by $ -
\frac{\chi}{2r^{2}} {\cal D}^{\ast}$, $\cal D$, and $\frac{-\eta^{2}}{2r^{2}}$
respectively (see Eqs.\ (8) and (9) of Ref. \cite{1}).

We assume the following general expression for the asymptotic solution of
$\psi_{\rm g}(r)$ at the spatial infinity\footnote{In fact, this is
essentially the same asymptotic form for $\psi_{\rm g}$ as obtained from
Eqs.\ (6), (23), and (44) of Ref.\ \cite{1}; see also the paragraph after
Eqs.\ (43) of same reference.}:
\begin{equation}
     \psi_{\rm g} \longrightarrow \sum_{n} A_{n} r^{n} e^{-i\omega r_{\ast}}
     + \sum_{m} B_{m} r^{m} e^{i\omega r_{\ast}},
\end{equation}
$A_{n}$ and $B_{m}$ are (complex) constant coefficients. Of course, the
summations in Eq.\ (12) start with the leading terms. Substituting (12) into
Eq.\ (2) we obtain
\begin{equation}
    \psi_{\rm g} \longrightarrow A r^{-1} e^{-i\omega r_{\ast}} +
    (B r^{3}  + B_{2} r^{2} + B_{1} r + B_{0} + B_{-1} r^{-1}) e^{i\omega
    r_{\ast}},
\end{equation}
where only the leading term of the form $e^{-i\omega r_{\ast}}$ is displayed
(we have omitted the subindices in the coefficients for the leading terms
for simplicity), since it is the only one involved in the calculations below.
For the terms of the form $e^{i\omega r_{\ast}}$, it is necessary to
consider the asymptotic series up to the fifth term, since this term gives
the leading contribution for the outgoing mode in $\Psi^{\rm B}_{0}$
\footnote{This is very easy to see, since $\overline{\Psi^{\rm B}_{0}}(r) =
{\cal{ \cal D }}^{4}\psi_{g}$, it is straightforward to demonstrate that
${\cal D}^{4}r^{m}e^{i\omega r_{\ast}} = (m-3)(m-2)(m-1)mr^{m-4}e^{i\omega
r_{\ast}}$, and then the leading contribution comes from the term  $m=-1$.
This fact is related with the so called ``peeling property" \cite{1}.},
which will be important for evaluating the constant $K_{\rm G}$. Equation\
(2) beside setting the leading terms, gives a {\it recurrence} relation for
the $A's$ and another for the $B's$. In particular, we can find that
\begin{equation}
      B_{-1} = \frac{B}{96\omega^{4}} \left| \frac{\eta^{2}}{2} (\eta^{2} +
      2) + 6 i\omega r_{+} \right|^{2},
\end{equation}
which will be the only relation that we shall need for our purpose. In a
fully similar way we can obtain, from Eq.\ (7), the following asymptotic
form of $\psi_{e}$ corresponding to electromagnetic perturbations:
\begin{equation}
    \psi_{\rm e} \longrightarrow a r^{-1} e^{-i\omega r_{\ast}} +
    (b r  + b_{0} + b_{-1} r^{-1}) e^{i\omega
    r_{\ast}},
\end{equation}
where $a$ and $b$ are the constant coefficients for the leading terms and
\begin{equation}
b_{-1}= -\frac{b}{8\omega^{2}}(\eta^{2}+2)^{2}.
\end{equation}

\begin{center}
{\uno B. At the horizon}
\end{center}
\vspace{1em}
For the asymptotic expansions  of the ingoing modes at the horizon, it is
convenient to introduce the new radial variable
\begin{equation}
y \equiv r-r_{+}, \qquad r_{+}=2M \qquad({\rm horizon\ of\ the\ black\
hole}),
\end{equation}
and then to perform an expansion of Eq.\ (2) and (7) at the leading order in
$1/y$, in a similar way to that performed at the leading orders in $r$ in
the previous section. We have for the gravitational case that
\begin{equation}
     \psi_{\rm g} \longrightarrow C y^{2} e^{-i\omega r_{\ast}},
\end{equation}
and
\begin{equation}
     \psi_{\rm e} \longrightarrow c y e^{-i\omega r_{\ast}},
\end{equation}
for the electromagnetic one; $C$ and $c$ are constant coefficients. \\[2em]

\noindent {\uno IV. CONSERVATION RELATIONS AND REFLECTION AND TRANSMISSION
COEFFICIENTS FOR THE ENERGY}
\vspace{1em} \\

The relevant quantities for wave amplitudes and energy fluxes can be
obtained directly from our potentials $\psi_{\rm G}$ and $\psi_{\rm E}$
(without involving the functions $X_{-k}$, $Y_{-k}$ and $Z_{k}$ of
Ref.\ \cite{1}) by noting from Eqs.\ (41) and (42) of Ref.\ \cite{1}
that
\begin{eqnarray}
     \overline{\Psi^{\rm B}_{0}} & \!\! = \!\! & [ {\cal D}^{4} \psi_{\rm g}
     (r)] e^{-i\omega t} {_{-2}}Y_{jm}, \nonumber \\
     \overline{\Psi^{\rm B}_{4}} & \!\! = \!\! & \frac{\eta^{2} (\eta^{2} +
     2)}{4r^{4}} \psi_{\rm g} (r) e^{-i\omega t} {_{2}}Y_{jm} -
     \frac{3iM\omega}{r^{4}} \overline{\psi_{\rm g} (r)} \, e^{i\omega t} \,
     \overline{{_{-2}}Y_{jm}},
\end{eqnarray}
for gravitational waves, and
\begin{eqnarray}
     \overline{\varphi^{\rm B}_{0}} & \!\! = \!\! & [ {\cal D}^{2}
     \psi_{\rm e}(r)] e^{-i\omega t} {_{-1}}Y_{jm}, \nonumber \\
     \overline{\varphi^{\rm B}_{2}} & \!\! = \!\! & \frac{(\eta^{2} +
     2)}{2r^{2}} \psi_{\rm e} (r) e^{-i\omega t} {_{1}}Y_{jm},
\end{eqnarray}
for electromagnetic waves.

In this manner, from Eqs.\ (13), (15), (20), and (21) one finds that at the
spatial infinity:
\begin{eqnarray}
     \frac{1}{4} r \overline{\Psi^{\rm B}_{0}} & \longrightarrow & \big[ 4
     \omega^{4} A e^{-i\omega (t + r_{\ast})} + \frac{6B_{-1}}{r^{4}}
     e^{-i\omega (t - r_{\ast})} \big] {_{-2}}Y_{jm}, \nonumber \\
     r \overline{\Psi^{\rm B}_{4}} & \longrightarrow & \frac{\eta^{2}
     (\eta^{2} + 2)}{4} B e^{-i\omega (t- r_{\ast})} {_{2}}Y_{jm} - 3 i
     M\omega \overline{B} e^{i\omega (t- r_{\ast})}\overline{{_{-2}}Y_{jm}},
     \nonumber \\
     \frac{1}{2} r \overline{\varphi^{\rm B}_{0}} & \longrightarrow & \big[
     -2 \omega^{2} a e^{-i\omega (t + r_{\ast})} + \frac{b_{-1}}{r^{2}}
     e^{-i\omega (t - r_{\ast})} \big] {_{-1}}Y_{jm}, \nonumber \\
     r \overline{\varphi^{\rm B}_{2}} & \longrightarrow & \frac{(\eta^{2} +
     2)}{2} b e^{-i\omega (t - r_{\ast})} {_{1}}Y_{jm}.
\end{eqnarray}
Similarly, from Eqs.\ (17)--(19), (20), and (21) one finds that, at the
horizon,
\begin{eqnarray}
     y^{2} \overline{\Psi^{\rm B}_{0}} & \longrightarrow & 4i\omega C r_{+}
     (1 - i\omega r_{+}) |1 + 2i\omega r_{+}|^{2} e^{-i\omega (t +
     r_{\ast})} {_{-2}}Y_{jm}, \nonumber \\
     y \overline{\varphi^{\rm B}_{0}} & \longrightarrow & -2i\omega c r_{+}
     (1 - 2i\omega r_{+}) e^{-i\omega (t + r_{\ast})} {_{-1}}Y_{jm}.
\end{eqnarray}
Note that at level of the quantities of Eqs.\ (22) and (23) for the waves
amplitudes, the description is essentially the same as given by the
quantities of Eqs.\ (47) and (48) of the Ref.\ \cite{1}; in fact, making a
comparison with those quantities we can find relations between the
corresponding coefficients:
\begin{eqnarray}
     4 \omega^{2} A & \!\! = \!\! & - [\eta^{2} (\eta^{2} + 2) \pm 12 i M
     \omega] A^{(\pm)}_{2}, \nonumber \\
     \frac{B}{4\omega^{2}} & \!\! = \!\! & - B^{(\pm)}_{2}, \nonumber \\
     2 \omega a & \!\! = \!\! & i \eta (\eta^{2} + 2) A^{(\pm)}_{1},
     \nonumber \\
     b & \!\! = \!\! & 2 i \omega\eta B^{(\pm)}_{1}, \nonumber \\
     r_{+} (1 - i\omega r_{+}) (1 - 2i\omega r_{+}) C & \!\! = \!\! &
     \frac{1}{2} [\eta^{2} (\eta^{2} + 2) \pm 12 i M \omega ] D^{(\pm)}_{2},
     \nonumber \\
     r_{+} (1 - 2i\omega r_{+}) c & \!\! = \!\! & \eta (\eta^{2} + 2)
     D^{(\pm)}_{1}.
\end{eqnarray}
On the other hand, from the fact that, in particular $K_{\rm G} |_{r_{\ast}
\rightarrow + \infty} = K_{\rm G} |_{r_{\ast} \rightarrow - \infty}$ (see
Eq.\ (6) and the paragraph at the end of Sec.\ II), from Eqs.\ (13), (18),
first of Eqs.\ (22) and (23) one finds that
\begin{equation}
     (3 B \overline{B_{-1}} - 2 \omega^{4} | A |^{2}) = -
     \frac{r^{2}_{+}}{2} | 1 + i\omega r_{+} |^{2} | 1 + 2 i\omega r_{+}
     |^{2} |C|^{2},
\end{equation}
where $B_{-1}$ is given in Eq.\ (14). Similarly, from Eqs.\ (10), (15),
(19), third of Eqs.\ (22), and second of Eqs.\ (23), one finds that, in the
electromagnetic case,
\begin{equation}
     2 (2\omega^{2} |a|^{2} + b \overline{b_{-1}}) = r^{2}_{+} | 1 + 2
     i\omega r_{+}|^{2} |c|^{2},
\end{equation}
where $b_{-1}$ is given in Eq.\ (16).

Furthermore, using the (limiting) formulae given in Eqs.\ (51)--(53), and
(56) in Ref.\ \cite{1} and our expressions (22) and (23), we find that
\begin{eqnarray}
     dE^{\rm grav}_{\rm in}/dt & \!\! = \!\! & \frac{4\omega^{6}}{\pi}
     |A|^{2}, \nonumber \\
     dE^{\rm em}_{\rm in}/dt & \!\! = \!\! & \frac{2\omega^{4}}{\pi} |a|^{2},
     \nonumber \\
     \langle dE^{\rm grav}_{\rm out}/dt \rangle & \!\! = \!\! & \frac{1}{64
     \pi\omega^{2}} [\eta^{4} (\eta^{2} + 2)^{2} + 144 M^{2} \omega^{2}
     |B|^{2}, \nonumber \\
     dE^{\rm em}_{\rm out}/dt & \!\! = \!\! & \frac{(\eta^{2} + 2)^{2}}{8\pi}
     |b|^{2}, \nonumber \\
     dE^{\rm grav}_{\rm hole}/dt & \!\! = \!\! & \frac{\omega^{2}}{\pi}
     r^{2}_{+} |1 + i\omega r_{+}|^{2} |1 + 2i\omega r_{+}|^{2} |C|^{2},
     \nonumber \\
     dE^{\rm em}_{\rm hole}/dt & \!\! = \!\! & \frac{\omega^{2}}{2\pi}
     r^{2}_{+} |1 + 2i\omega r_{+}|^{2} |c|^{2}.
\end{eqnarray}

Thus, from Eqs.\ (25)--(27) we obtain that
\begin{eqnarray}
     dE^{\rm grav}_{\rm in}/dt & \!\! = \!\! & \langle dE^{\rm
     grav}_{\rm out}/dt \rangle + d E^{\rm grav}_{\rm hole}/dt, \nonumber \\
     d E^{\rm em}_{\rm in}/dt & \!\! = \!\! & d E^{\rm em}_{\rm out}/dt + d
     E^{\rm em}_{\rm hole}/dt,
\end{eqnarray}
which show explicity that the energy carried by the gravitational or the
electromagnetic perturbations is conserved. We point out that, the
``conservation relations" (25) and (26) (such as those in Ref.\ \cite{1})
can be properly interpreted as expressions of the conservation of energy
only after the energy fluxes (27) are derived. In this manner, the original
continuity equations (4) and (9) for non-hermitian systems lead, finally, to
a conserved quantity, the energy for the field perturbations. Furthermore,
from expressions (27) one finds that
\begin{eqnarray}
    \frac{\langle dE^{\rm grav}_{\rm out}/dt \rangle}{dE^{\rm grav}_{\rm in}/dt}
     & \!\! = \!\! & \frac{[\eta^{4}(\eta^{2}+ 2)^{2} + 144 M^{2}
     \omega^{2}]}{256 \omega^{8}} \frac{|B|^{2}}{|A|^{2}} =
     \frac{|B^{(\pm)}_{2}|^{2}}{|A^{(\pm)}_{2}|^{2}}, \nonumber \\
     \frac{ dE^{\rm grav}_{\rm hole}/dt }{ dE^{\rm grav}_{\rm in}/dt} & \!\! = \!\! &
     \frac{r^{2}_{+} |1 + i\omega r_{+}|^{2} |1 + 2 i\omega
     r_{+}|^{2}}{4\omega^{4}} \frac{|C|^{2}}{|A|^{2}} = \frac{|D^{(\pm)}_{2}
     |^{2}}{|A^{(\pm)}_{2}|^{2}}, \nonumber \\
     \frac{dE^{\rm em}_{\rm out}/dt}{dE^{\rm em}_{\rm in}/dt} & \!\! = \!\! &
     \frac{(\eta^{2}+ 2)^{2}}{16 \omega^{4}} \frac{|b|^{2}}{|a|^{2}}
     = \frac{|B_{1}|^{2}}{|A_{1}|^{2}}, \nonumber \\
     \frac{dE^{\rm em}_{\rm hole}/dt}{dE^{\rm em}_{\rm in}/dt} & \!\! = \!\! &
     \frac{r^{2}_{+} |1 + 2i\omega r_{+}|^{2}}{4 \omega^{2}}
     \frac{|c|^{2}}{|a|^{2}} = \frac{|D_{1}|^{2}}{|A_{1}|^{2}},
\end{eqnarray}
where the second equalities follow from the relations (24). Thus, our energy
reflection and transmission coefficients for the field perturbations
coincide with those in Ref.\ \cite{1} given in terms of the reflection and
transmission coefficients for the potential barriers appearing in the
Schr\"odinger-type equations. However, it is worth to point out that, in the
present approach, it has not been mentioned explicity the existence of some
``efective potential barriers". Furthermore, in this approach, there exists,
in general, only one reflection or transmission coefficient for each field
perturbation, avoiding the ``redundance" which appears in the traditional
approach because the existence of various potential barriers.

The polarization changes of the field perturbations are analyzed at the
level of Eqs.\ (47) and (48) in Ref.\ \cite{1}, therefore, the conclusions
of that reference are also valid in our present approach (see paragraph
after Eq.\ (23)). \\[2em]

\begin{center}
{\uno V. CONCLUDING REMARKS}
\end{center}
\vspace{1em}
In conclusion, we have demonstrated that one can obtain essentially the same
physical results, starting from a {\it manifestly} non-hermitian system,
to those obtained from the traditional approach via (hermitian)
Schr\"odinger-type equations. This suggests that a fully similar analysis
can be performed in the perturbations of exact solutions in the setting of
the modern theories, where, as pointed out in Sec.\ I, non-hermitian systems
appear. In particular those solutions corresponding to black holes in string
theory, for which some results analogous to those traditionally known for
classical black holes have not established still. Works along these lines
are in progress and will be the subject of forthcoming communications.

On the other hand, although the result on the existence of conserved
currents in Sec.\ II has been established assuming only tensor fields and
the presence of a single equation, such a result can be extended in a direct
way to equations involving spinor fields, matrix fields, and the presence of
more than one field \cite{6}. The possible applications of this very general
result in the theories involving gravity and matter fields (and other areas
of physics) are open questions. \\[2em]

\begin{center}
{\uno ACKNOWLEDGMENT}
\end{center}
\vspace{1em}

 The author wants to thank Dr. G. F. Torres del Castillo for his
comments on the manuscript and suggestions. The author would like to thank
Professor R. M. Wald for the kind hospitality provided at The Enrico Fermi
Institute, University of Chicago. The author acknowledges the financial
support received from CONACYT and the Sistema Nacional de Investigadores.
\\[2em]


\begin{thebibliography}{}
\setlength{\itemsep}{-.50em}
\bibitem{1} G. F. Torres del Castillo, J.\ Math.\ Phys.\ {\bf 37}, 5684
(1996).
\bibitem{2} P. K. Silaev and S. G. Turyshev, Gen.\ Rel.\ Grav.\ {\bf 29},
417 (1997).
\bibitem{3} R. Cartas-Fuentevilla, Phys.\ Rev.\ D {\bf 57}, 3443 (1998).
\bibitem{4} G. F. Torres del Castillo and J. C. Flores-Urbina, 
  Gen.\ Rel.\ Grav.\ {\bf 31}, 1315 (1999).
\bibitem{5} R. M. Wald, Phys.\ Rev.\ Lett.\ {\bf 41}, 203 (1978).
\bibitem{6} G. F. Torres del Castillo, Gen.\ Rel.\ Grav.\ {\bf 22}, 1085
(1990).
\bibitem{7} G. A. Burnett and R. M. Wald, Proc.\ R.\ Soc.\ London {\bf A430},
57 (1990).
\bibitem{8} J. Lee and R. M. Wald, J.\ Math.\ Phys.\ {\bf 31}, 3 (1990).
\bibitem{9} R. M. Wald, Proc.\ R.\ Soc.\ A{\bf 369}, 67 (1979).

\end{thebibliography}
\end{document}